\documentclass[useAMS,usenatbib,usegraphicx,referee,a4paper]{mn2e}
\newcommand {\aplt} {{\raise-.5ex\hbox{$\buildrel<\over\sim$}}} 
\usepackage{amsmath}
\usepackage{url}

 \title[Water-depend colour terms in space-based infrared surveys]{Detecting free-floating planets using water-depend colour terms in the next generation of infrared space-based surveys}

   \author[N.R.\ Deacon et al.]{N.R. Deacon$^{1,2}$ \\
   $^1$Max Planck Institute for Astronomy, Konigstuhl 17, Heidelberg, 69117, Germany\\
$^2$Centre for Astrophysics Research, University of Hertfordshire, College Lane, Hatfield, AL10 9AB, UK\\
             }
 
\begin{document}
 \date{}
 \pagerange{\pageref{firstpage}--\pageref{lastpage}} \pubyear{2015}
 \maketitle
 \label{firstpage}
  \begin{abstract}
The next decade will see two large-scale space-based near-infrared surveys, Euclid and WFIRST. This paper shows that the subtle differences between the filters proposed for these surveys and those from ground-based photometric systems will produce a ground-space colour term that is dependent on water absorption in the spectra of astronomical objects. This colour term can be used to identify free-floating planets in star forming regions, mimicing a successful ground-based technique that uses a filter sensitive to water absorption. This paper shows that this colour term is an effective discriminant between reddened background stars and ultracool dwarfs. This represents just one science justification for a Galactic Plane survey in the event of an extension to the Euclid mission beyond its original timeframe.
  \end{abstract}
%

\section{Introduction}
Since the end of the 1990s, near-infrared surveys have transformed our view of the universe, from high-redshift quasars \citep{Mortlock2011} and early galaxy formation \citep{Foucaud2007,Bouwens2015,Finkelstein2015} to ultracool dwarfs in our own Galactic back-yard \citep{Burgasser2002,Burningham2011}. To date, most wide-field surveys short of two and a half microns have been conducted from the ground. Hence surveys such as 2MASS \citep{Skrutskie2006}, UKIDSS \citep{Lawrence2007} and VISTA\footnote{http://casu.ast.cam.ac.uk/surveys-projects/vista/technical/filter-set} use filter sets which avoid wavelengths dominated by telluric water absorption. The next decade will see two large infrared surveys of the sky conducted from space. The Euclid NISP instrument \citep{Laureijs2011} will map one third to one half of the sky in $Y$, $J$ and $H$ with the goal of measuring the acceleration of the universe. WFIRST \citep{Spergel2015}, which focusses both on measuring cosmic acceleration and detecting extrasolar planets, will use similar near-infrared filters to survey large areas of the sky.

Ultracool dwarfs are objects with spectral types later than M7. These are known to have significant water absorption in their atmospheres at similar wavelengths to telluric water absorption. Indeed indices based on water absorption provide a useful tool for determining the spectral classification of late M \citep{Allers2013}, L \citep{McLean2003,Allers2013} and T \citep{Burgasser2006} dwarfs. L dwarfs have red colours in the near-infrared, most probably due to dust clouds in their photospheres; whilst the cooler T dwarfs have blue infrared colours due to methane absorption and a lack of photospheric dust clouds. The transition between these two spectral classes produces a rapid change in spectral type and is often associated with photometric variability \citep{Radigan2014a}. 

Substellar objects lack a stable internal energy source from hydrogen fusion. Thus, unlike main sequence stars there is a degeneracy between mass and age. Free-floating planetary mass objects have masses under the deuterium-burning dividing line between planets and brown dwarfs and yet move through space without a parent star. These can exist both as field objects \citep{Liu2013} and as members of young star forming regions \citep{AlvesdeOliveira2012}. At young ages, giant free-floating planets can be warm enough to have M, L and T spectral types, similar to many field brown dwarfs.These free-floating planets have lower observed surface gravity than field brown dwarfs and (based on their ages and temperatures) are predicted to have masses below the deuterium-burning limit by theoretical models \cite{Saumon2008}.

Star forming regions are often found within regions of the sky with high amounts of background reddening. This means that photometrically it is hard to distinguish red L type free-floating planetary mass objects from highly reddened background stars. \cite{Allers2007} present a novel method to solve this problem. They proposed a specialist $W$ band filter sandwiched between the ground-based $J$ and $H$ bands at around 1.4 microns. M, L and T dwarfs have significant water absorption around this wavelength with methane also contributing to absorption in T dwarfs. Reddened background stars will be dominated by continuum emission in this wavelength range. As this filter covers a wavelength range with telluric water absorption it is best used from the driest, most stable sites. Recently a $W$ band filter has been installed on CFHT on Mauna Kea. 

Ground-based near-infrared filter sets avoid wavelength regions with significant telluric water absorption, leaving gaps between the $J$ and $H$ and $H$ and $K$ filters (see for example the MKO photometric system \citealt{Tokunaga2002}). Space-based surveys such as Euclid and WFIRST are not affected by telluric water absorption. Hence there is no need for their near-infrared filters to avoid wavelengths where water absorption occurs. This means that space-based filters will sample spectral regions where ultracool dwarfs have significant water absorption whilst ground-based filters will not. This paper shows that this difference in wavelength coverage leads to near-infrared colour terms between ground-based and space-based filter, mimicing the effect of observations with a specialist $W$ band filter. 

In this paper synthetic photometry is used to determine that a combination of space-based and ground-based filters gives a colour term that distinguishes L-type objects from reddened stars. A short discussion then follows on how this technique could be used in both the planned Euclid and WFIRST surveys and how future surveys with these missions could be informed by this technique.

\section{Synthetic Photometry}
\begin{figure}
 \setlength{\unitlength}{1mm}
 \begin{picture}(100,100)
 \includegraphics{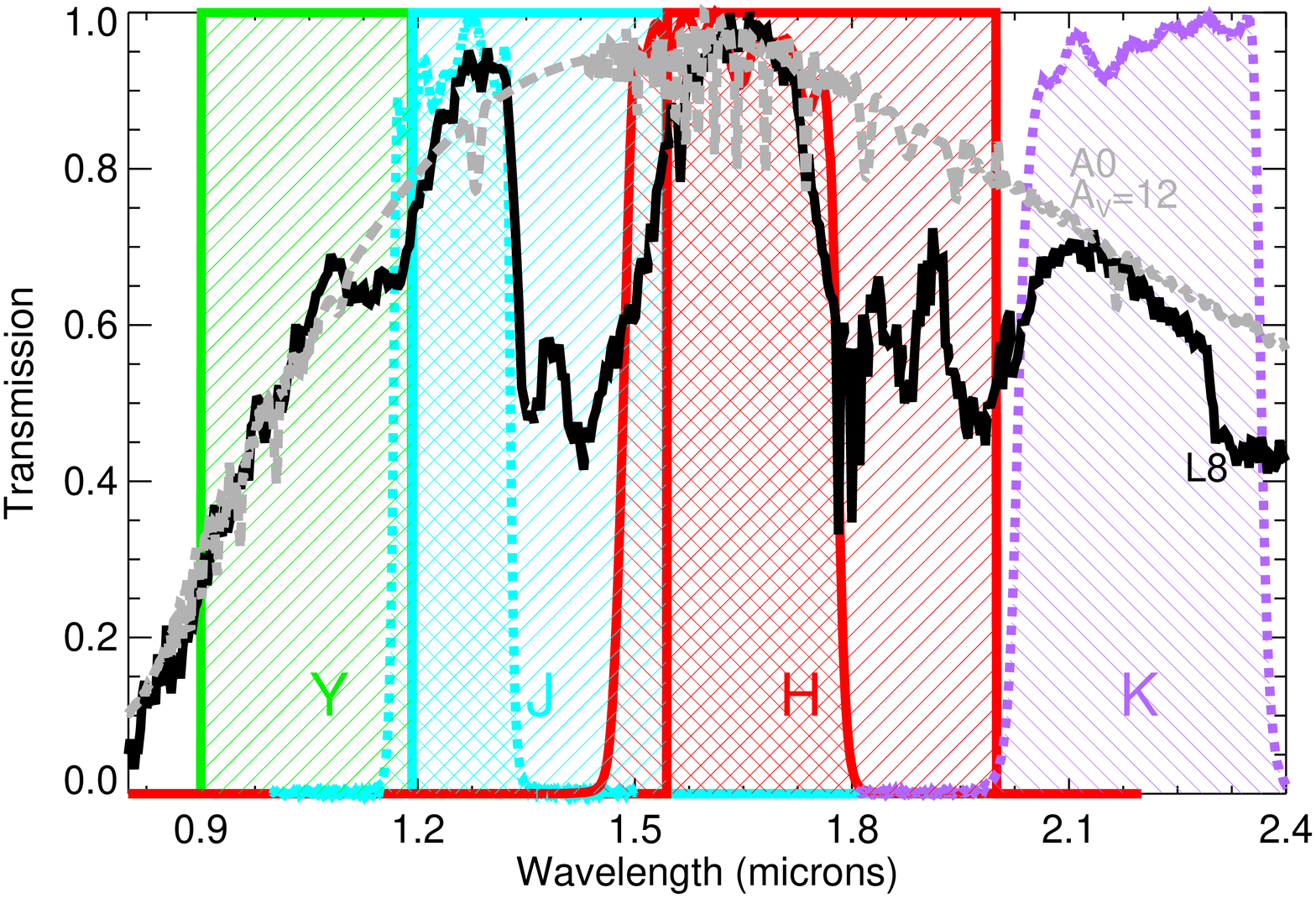}
    \end{picture}
   \caption{The filter response curves for MKO \protect\citep{Tokunaga2002} $J$, $H$ and $K$ (dotted coloured lines) and Euclid $Y$, $J$ and $H$ (solid coloured lines). The solid black line shows the L8 standard \protect\citep{Kirkpatrick1999,Burgasser2007} and the dashed grey line an A0 template from \protect\cite{Pickles1998a} which has been reddened. Detector response is not included in the curves plotted here and all filters have been normalised to have a maximum response of 1.0 to make the plot clearer}
              \label{Filt_euclid}%
    \end{figure}

\begin{figure}
 \setlength{\unitlength}{1mm}
 \begin{picture}(100,100)
 \includegraphics{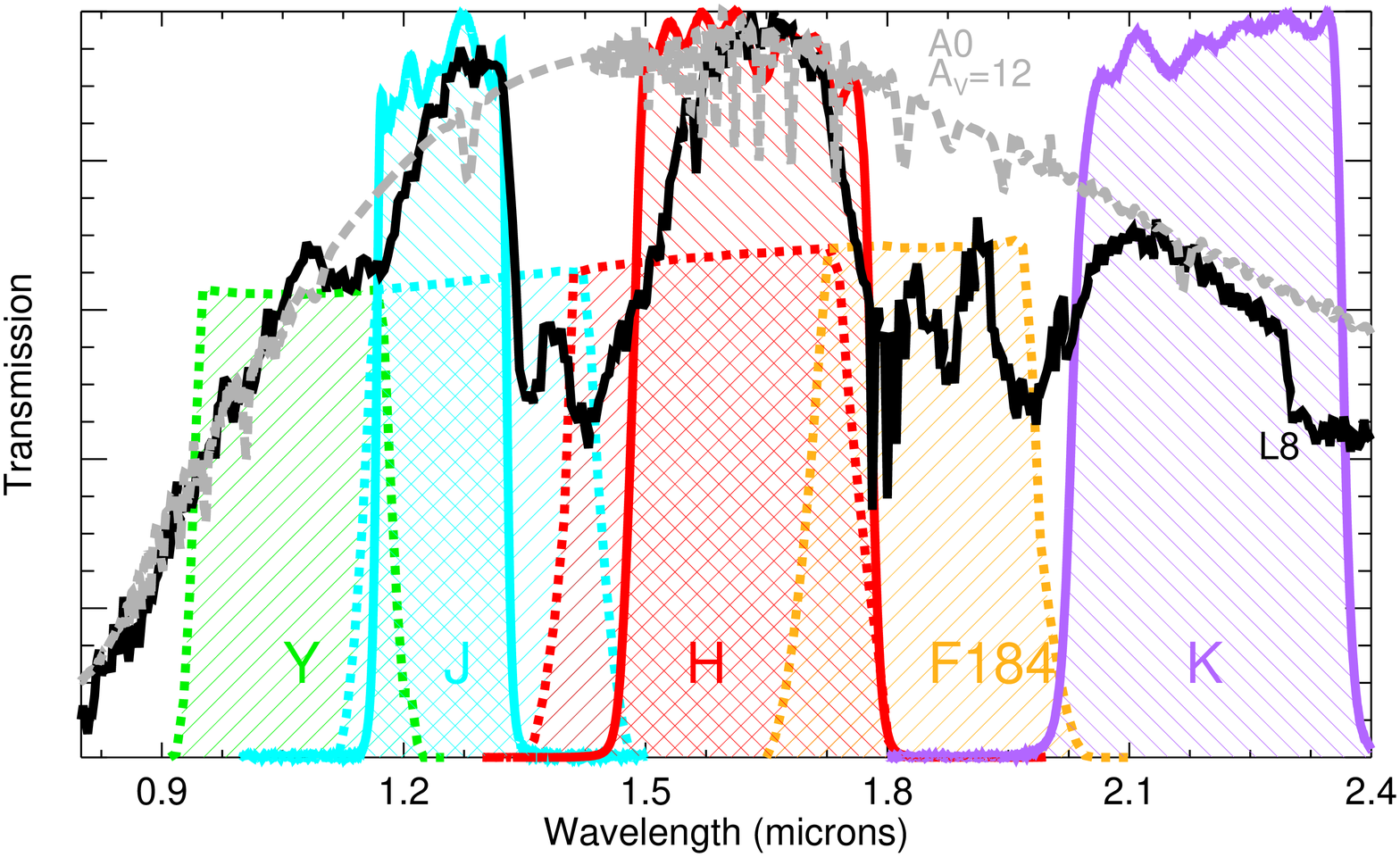}
    \end{picture}
   \caption{The filter response curves for MKO \protect\citep{Tokunaga2002} $J$, $H$ and $K$ (dotted coloured lines) and WFIRST $Y$, $J$, $H$ and $F184$ (solid coloured lines). The solid black line shows the L8 standard \protect\citep{Kirkpatrick1999,Burgasser2007} and the dashed grey line an A0 template from \protect\cite{Pickles1998a} which has been reddened. Detector response is not included in the curves plotted here and MKO filters have been normalised to have a maximum response of 1.0 to make the plot clearer}
              \label{Filt_wfirst}%
    \end{figure}

To test the effectiveness of combine space-based and ground-based data, synthetic photometry was calculated for both ultracool dwarfs and reddened field stars. For Euclid photometry top-hat filter profiles were produced using the limits from the current Euclid-NISP website\footnote{\url{https://www.euclid-ec.org/?page_id=2490}}. This approach is similar to \cite{Holwerda2018}. For WFIRST the filter and optical system response curves were taken from the WFIRST Reference Information Cycle 7\footnote{\url{https://wfirst.gsfc.nasa.gov/science/WFIRST_Reference_Information.html}}. The filter curves for Euclid are shown in Figure~\ref{Filt_euclid} and those for WFIRST are shown in Figure~\ref{Filt_wfirst}. Both figures include the spectrum of the L8 standard \citep{Kirkpatrick1999,Burgasser2007}. Note how the $J$ and $H$ bands for the two missions cover both the peaks in emission from the L dwarf and the two water absorption bands around 1.4 and 1.9 microns. By contrast the ground-based MKO system \citep{Tokunaga2002} avoids these water bands and only covers the parts of the L8 spectrum with significant emission. A magnitude is a measurement of average flux density across a passband. As the flux of an L8 dwarf is significantly denser in the MKO $J$ band filter than in the Euclid and WFIRST $J$ band filters, it is likely there will be a significant, water band dependent, colour term between the two.

The synthetic photometry was calculated using spectra of ultracool dwarfs was drawn from the SpeX prism library\footnote{\url{http://pono.ucsd.edu/~adam/browndwarfs/spexprism}} including M dwarf standards from \cite{Kirkpatrick2010}, L dwarf standards from \cite{Kirkpatrick1999} and T dwarf standards from \cite{Burgasser2006}. For reddened stars the A0 and M0III templates from \cite{Pickles1998a} and the reddening law from \cite{Cardelli1989} applied in the {\tt ccm\_UNRED IDL} routine were used. Reddening values ranging from $A_V=0$ to $A_V=20$ were applied.

\begin{figure}
 \setlength{\unitlength}{1mm}
 \begin{picture}(100,50)
 \includegraphics{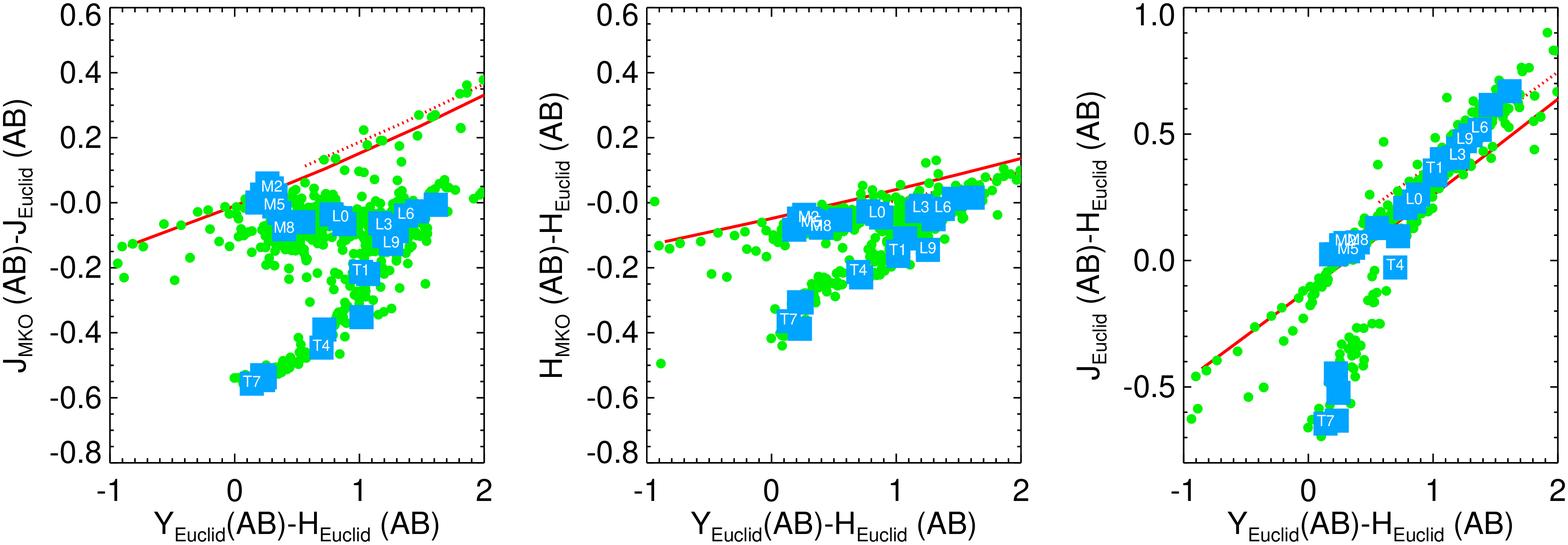}
    \end{picture}
   \caption{The infrared Euclid colours of known ultracool dwarfs (green) and M, L and T standards (blue squares). Every third subtype is marked on the corresponding blue square to show the general trend. Also plotted are the reddening lines for an A0 dwarf (solid red line) and an M0 giant (dashed line). A clear, spectral-type dependent colour term can be seen in both $J$ and $H$.}
              \label{mko_euclid}%
    \end{figure}

Figure~\ref{mko_euclid} shows the colour terms between Euclid and MKO photometry sets. Using only Euclid data (right-hand panel) the L~dwarfs lie along the reddening line. The ultracool dwarf sequence only deviates significantly from the reddening line for spectral types later than T4. By contrast when combined with MKO data, Euclid photometry can be used to separate reddened stars from ultracool dwarfs. The left-hand panel of Figure~\ref{mko_euclid} shows a $J_{MKO}-J_{Euclid}$ colour term for L dwarfs, separated from reddened stars by at least 0.2 magnitudes with better separation at later spectral types. 

\begin{figure}
 \setlength{\unitlength}{1mm}
 \begin{picture}(100,50)
 \includegraphics{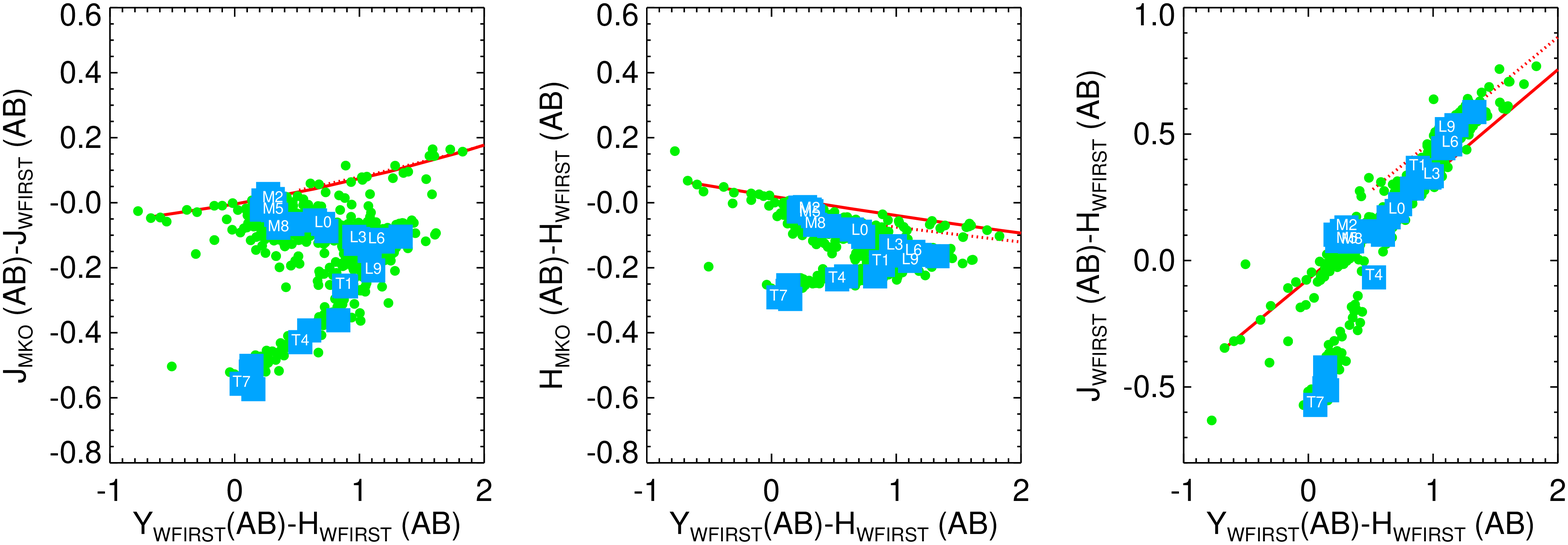}
    \end{picture}
   \caption{The infrared WFIRST colours of known ultracool dwarfs (green) and M, L and T standards (blue squares). Every third subtype is marked on the corresponding blue square to show the general trend. Also plotted are the reddening lines for an A0 dwarf (solid red line) and an M0 giant (dashed line). A clear, spectral-type dependent colour term can be seen in both $J$ and $H$.}
              \label{mko_wfirst}%
    \end{figure}

Figure~\ref{mko_wfirst} shows the colour terms between the WFIRST and MKO filter sets. There is a similar colour term to the Euclid-MKO comparison in the $J$ band. However it is worth noting that the smaller $H$ band colour term is subtly different to the Euclid-MKO $H$ band colour term. This is due to the WFIRST $H$ band sampling a significantly bluer wavelength range than the Euclid $H$ band.

\begin{figure}
 \setlength{\unitlength}{1mm}
 \begin{picture}(100,50)
 \includegraphics{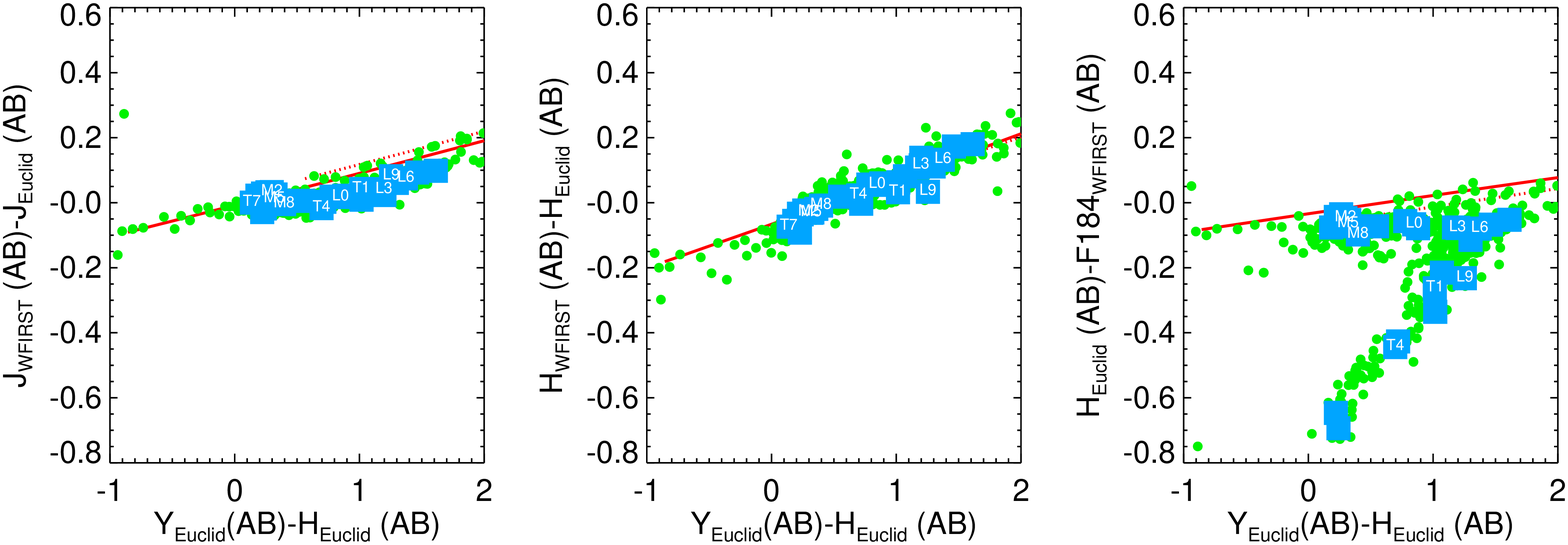}
    \end{picture}
   \caption{The infrared WFIRST and Euclid colour combinations of known ultracool dwarfs (green) and M, L and T standards (blue squares). Every third subtype is marked on the corresponding blue square to show the general trend. Also plotted are the reddening lines for an A0 dwarf (solid red line) and an M0 giant (dashed line). No colour term between the two surveys appears to be useful, only a small effect is seen in the $H_{Euclid}-F184_{WFIRST}$ combination.}
              \label{wfirst_euclid}%
    \end{figure}

Figure~\ref{wfirst_euclid} shows the colour terms between the WFIRST and Euclid filter sets. No strong colour terms are found. There is a small colour term between $H_{Euclid}$ and $F184_{WFIRST}$. This is likely due to the Euclid and WFIRST filters being wider and always including both emission peaks and absorption bands in L dwarf spectra. The MKO $J$-band filter by contrast contains only the J-band emission peak in an ultracool dwarf spectrum.

\section{Discussion}
Euclid is a cosmology mission aiming to measure the equation of state of dark energy. Hence its planned survey area is primarily extragalactic covering 15,000 sq. deg. at high Galactic and Ecliptic latitudes\footnote{\url{https://www.euclid-ec.org/?page_id=2581}}. This will limit its usefulness for studying cool objects in star forming regions. However if the Euclid spacecraft is operational beyond its scheduled 5 year mission there may be the possibility to undertake additional surveys. Identifying free-floating planets in star forming regions would be just one of a number of science cases for a Euclid Galactic survey.

Similarly it is planned that WFIRST will undertake a high latitude survey to study dark energy with an additional Galactic Bulge survey for microlensing science. It is however also proposed that WFIRST would have a guest observer programme, leaving open the possiblity of studying star forming regions in the Galactic Plane. 

Detecting ultracool dwarfs by a space-ground near-infrared colour term requires deep ground-based data. The most effective filter for such companion observations will be the $J$-band as this has the largest colour term when compared to both Euclid and WFIRST. Large ground-based surveys such as the UKIRT Hemisphere Survey \citep{Dye2018} and the VISTA Hemisphere Survey \citep{McMahon2013} will cover over 90\% of the sky with $J$-band data to $J_{AB}\sim$20.5\,mag. This is 3.5 magnitudes shallower than the planned Euclid survey \citep{Laureijs2011}. While some star forming regions may be covered by deeper legacy data from previous surveys for late-type members, achieving comparable depths to the Euclid extragalactic survey will require dedicated observations on 8\,m-class telescopes. 

Ground-based near-infrared surveys have also revolutionised studies of galaxy formation (\citep{Foucaud2007}. Such extragalactic surveys consider ultracool dwarfs to be contaminants. Pre-existing ground-based observations to comparable depth to Euclid survey data could be used in combination with new space-based data to identify late-type substellar interlopers in galaxy samples.

PSO~J318.5$-$22 \citep{Liu2013} is a late-L free-floating planet in the $\beta$~Pictoris moving group at a distance of 24.6$\pm$1.4\,pc \citep{Liu2013}. Such objects can be detected in the field but as most of the sky does not have large amounts of background reddening, contamination from highly reddened background stars is rare for field samples of L dwarfs outside the Galactic Plane. It is in star forming regions where both young, free-floating planets and contamination from reddened background stars combine to make water-dependent colour terms a useful technique. A Euclid survey in the Galactic Plane and star forming regions at low Galactic latitude to the same $J_{AB}\sim24$\,mag. depth as the main Euclid survey would detect free-floating planets with the same absolute $J$ band magnitude as PSO~J318.5$-$22 in every star forming region within 300\,pc to a significance of 15$\sigma$. As these regions (such as the Sco-Cen complex, Chameleon and Taurus) are younger than $\beta$~Pictoris such analogues would have lower masses than PSO~J318.5$-$22.

The NIRCam instrument on the {\it James Webb Space Telescope} (JWST \citealt{Horner2004}) will include an F140M filter which will cover the water absorption band around 1.4 microns. This will allow water imaging of nearby star forming regions. However JWST will be an extremely competitive telescope with a field of view $\sim$200 times smaller than Euclid. These factors will limit the star forming regions which will be sampled by NIRCam. JWST's excellent near and mid-infrared spectroscopy abilities will make it ideal for following-up any free-floating planets discovered using a Euclid or WFIRST ground-space water absorption colour term.

The technique outlined here is most effective to identify objects with broad absorption features between the $J$ and $H$ bands. As shown in Figure~\ref{Filt_wfirst} the WFIRST filter system includes an $H$ band which cuts off from 1.7 to 1.8 microns and an $F184$ filter which is sensitive to emission beyond 1.65 microns. This latter filter covers the water and methane absorption in objects cooler than 1400\,K. Hence such objects will have a very blue $H-F184$ colour with the coldest field brown dwarfs and free-floating planets being $F184$ dropouts.

The technique outlined here (like that of \citealt{Allers2007} is only the first step in identifying if an object is a free-floating planet or brown dwarf. Follow-up spectroscopic observations would be needed to determine if the object does indeed have L or T spectral type and in the case of free-floating planets if they have low gravity spectral signatures. The spectral indices of \cite{Allers2013} would allow true free-floating planet in a star forming regions to be disentangled from rare interloping foreground brown dwarfs.

Finally it should be noted that specially manufactured $W$-band filters only cover the wavelength range where water absorbs. This means that objects with water absorption will have a larger colour colour difference between a $W$-band filter and a broadband filter than between space-based and ground-based broadband filter sets. This means specialist surveys undertaken in the next few years with $W$-band will be able to better discriminate between reddened stars and ultracool dwarfs than using space-based colour terms, albeit to shallower depths than Euclid of WFIRST.


\section{Conclusions}
This work shows that the different passbands of ground-based and proposed space-based photometric systems produce a colour term dependent on water absorption in astronomical objects. In a decade it will be possible to use this colour term to mimic the powerful water imaging technique for detecting young free-floating planets. Euclid and WFIRST will concentrate on surveys at high galactic latitude for their core science goals. This technique is one possible justification for future Galactic plane surveys using these spacecraft. 
\section*{Acknowledgments}
The author would like to thank Katelyn Allers and Beth Biller for helpful discussions on this idea presented in this paper and Knud Jahnke for helpful discussions on Euclid/NISP.
This research has benefitted from the SpeX Prism Spectral Libraries, maintained by Adam Burgasser at \url{http://pono.ucsd.edu/~adam/browndwarfs/spexprism}. The author acknowledges acknowledges the support of the DFG priority program SPP 1992 "Exploring the Diversity of Extrasolar Planets (Characterising the population of wide orbit exoplanets)"
\bibliographystyle{mn2e} 
 \bibliography{../ndeacon} 
\end{document}